\documentstyle[12pt]{article}
\begin{document}

\baselineskip=14pt plus 0.2pt minus 0.2pt
\lineskip=14pt plus 0.2pt minus 0.2pt

\begin{flushright}
hep-ex/9506015\\
LA-UR-95-1604\\
\end{flushright}

\begin{center}
\Large{\bf \footnote{
Email:  mhh@lanl.gov; feng@dfi.aau.dk; goldman@hotelcal.lanl.gov;
king@p15mv1.lanl.gov; rxl@lepsa.phys.psu.edu; mmn@pion.lanl.gov;
smith@lepsa.phys.psu.edu}ARE ANTIPROTONS FOREVER?}

\vspace{0.25in}

\large

\bigskip

M. H.\ Holzscheiter,$^a$ X.\ Feng,$^b$ T.\ Goldman,$^c$ N. S.
P.\ King,$^a$ \\ R. A.\ Lewis,$^d$  M. M.\ Nieto$^{ce}$ and  G.
A.\ Smith,$^d$

\end{center}

\begin{flushleft}

\small

$^a${\it Physics Division, Los Alamos National Laboratory, Los Alamos,
NM 87545, U.S.A.} \\
$^b${\it University of Aarhus, Institute for Physics and Astronomy,
8000 Aarhus-C, DENMARK} \\
$^c${\it Theoretical  Division, Los Alamos National Laboratory,
Los Alamos, NM 87545, U.S.A.} \\
$^d${\it Pennsylvania State University, University Park, PA 16802, U.S.A.} \\
$^e${\it Abteilung f\"ur Quantenphysik, Universit\"at Ulm, D-89069 Ulm,
GERMANY}

\end{flushleft}

\normalsize

\vspace{0.2in}
\begin{center}
{ABSTRACT}

\end{center}

\begin{quotation}

\baselineskip=0.333in

Up to one million antiprotons from a single LEAR spill have been
captured in a large Penning trap.  Surprisingly, when the antiprotons
are cooled to energies significantly below 1 eV, the  annihilation rate
falls below background.  Thus, very  long storage times for antiprotons
have been demonstrated in the trap, even at the compromised vacuum
conditions imposed by the experimental set up.  The significance for
future ultra-low energy experiments, including portable antiproton
traps, is discussed.


\vspace{0.25in}

\end{quotation}

\vspace{0.3in}

\newpage

\baselineskip=0.333in

An experiment to measure the gravitational acceleration of antiprotons
 is under preparation at the Low Energy Antiproton Ring (LEAR) at CERN
\cite{PS200}. The experiment proposes to use a time-of-flight technique
\cite{gn}, as pioneered in an experiment which measured the
gravitational acceleration of electrons \cite{wf}. A critical
requirement for such an experiment is  a sufficiently large number of
antiprotons at sub-eV energies in order to assemble a time-of-flight
spectrum with sufficient statistics.

The lowest-energy antiprotons  currently available are produced at
LEAR. Here antiprotons are delivered at energies as low as $5.9$ MeV.
A gap of at least $10$ orders of magnitude in energy has to be bridged
before a meaningful measurement of the gravitational acceleration of
antiprotons can be attempted.

To achieve this energy reduction we have developed a large Penning trap
system which is matched to the output phase space of the LEAR facility.
An antiproton bunch of $200$ ns duration, containing up to $10^9$
antiprotons, is transmitted through a thin foil in which the energy of
the individual particles is reduced by multiple collisions.  With a
properly chosen foil thickness up to $0.6$ \% of the incident
antiprotons  emerge from the foil with less than $12.5$ keV kinetic
energy.

These particles are dynamically captured in the Penning trap by rapidly
switching the entrance electrode potential while the bunch is inside
the trap volume. Once captured, the antiprotons are cooled by an
electron cloud which has been stored in the trap in preparation for the
capture. During recent tests of this system we have succeeded in the
capture of up to one million antiprotons from a single bunch from LEAR.
Up to $65$ \% of the captured particles were cooled to sub-eV energies
and collected in a $1$ cm$^3$ region at the center of the trap.

Using a set of scintillators mounted externally to the vacuum system we
are able to monitor the annihilation of the antiprotons on the residual
gas molecules  during the cool-down period.  When all particles have
been collected in the central well and have been cooled
 below $1$ eV, no annihilation can be observed above the ambient
background of approximately 1-2 counts per second.

This result is, at first glance, in contradiction to what one would
expect to happen since the  annihilation cross section at low energy is
generally assumed to have a $1/v$ dependence. As a result of this
effect, antiprotons were stored for significantly long periods of time,
even though the residual gas pressure in the system was estimated to be
equal to or greater than $10^{-11}$  Torr.  Note that our result is of
different origin than the long storage times obtained by the PS196
collaboration \cite{fei}. There a fully cryogenic vacuum  system was
used.  Their long storage time  was simply attributed to an
extremely-low residual gas density.  Effects discussed here were not
considered.

We now describe our results in detail and comment on their
significance.  Charged particles may be confined in vacuum by a
superposition of an electric quadrupole field and a strong, axial,
magnetic field, a combination typically referred to as a Penning trap
\cite{pt}.  One needs to ensure that all the antiprotons emerging from
the degrading foil during a single LEAR pulse and having a kinetic
energy of less than $12.5$ keV are still within the trap volume when
the potential at the entrance electrode is ramped up.  This requires an
axial dimension of the trap of  about $50$ cm.

To meet this requirement we have constructed an `open-end-cap' Penning
trap \cite{Gab_o_e}.  It contains  5 cylindrical electrodes of inside
diameters 2.8 cm and with other dimensions carefully chosen to form a
harmonic potential at the center.  Additionally there are two
high-voltage electrodes, located at the entrance and the exit of the
trap. The entrance electrode consists of a $5$ mil, gold-coated,
aluminum foil of diameter 0.6 cm, which also serves as the degrading
foil.  The exit electrode was chosen to be an open cylinder (of inside
diameter 2.8 cm) to allow  ejection of the antiprotons from the trap
subsequent to their capture and cooling.

This trap is located in the bore of a superconducting magnet capable of
producing an axial magnetic field of up to $6$ Tesla.  Figure 1
displays a schematic lay-out of the entire set-up, including the
location of the external scintillators used to monitor antiproton
annihilations.

The following is a brief description of a normal measurement cycle.
The central, harmonic well of the trap is preloaded with typically
$10^9$ electrons from an electron gun located in the fringe field of
the magnet.  These electrons  quickly cool by synchrotron radiation to
equilibrium with the ambient temperature of the system ($\approx 10$
K). Initially the entrance foil potential is held at ground while the
exit electrode is at full potential. Antiprotons  from LEAR traverse
the beam profile monitor, generating a trigger  for  the high-voltage
switch to  the entrance foil. The antiprotons are slowed down in the
foil. Those emerging from it at kinetic energies below the exit
electrode potential are reflected back towards the entrance. The
potential at the entrance electrode is ramped up to the desired
potential in less than $100$ ns by a commercial switch \cite{Behlke}.
This captures the antiprotons in the $50$ cm long (non-harmonic) well
of this ``catching trap." Due to scattering on the cold electrons the
antiprotons lose energy and eventually collect in the inner, harmonic
region of the trap.

Antiprotons stored in either the the long, non-harmonic well or the
inner, harmonic well of the trap can be detected by lowering the
respective electrical potentials.  Escaping antiprotons  will follow
the magnetic field lines, strike the surface of the down-stream
radiation baffle, and annihilate.  External scintillators S5-S8 (see
figure 1) detect the annihilations and the information is stored in a
multichannel analyzer.  If the time constant for reducing the potential
is chosen to be much longer than the oscillation period in the trap,
the resulting `time of arrival' spectrum directly reflects the energy
distribution of the particles in the trap before the release.

In Figure 2 we show the total number of antiprotons detected in the
inner, harmonic well (normalized to the number of antiprotons initially
captured) vs. the cooling time. We find that after approximately $600$
seconds as much as $65$ \% of the initially captured antiprotons were
cooled into the inner well. The solid line shows the result of a fit to
a cooling time constant of $175$ sec.

Figure 3 shows the energy spectrum of those antiprotons released from
the inner, harmonic trap after $1500$ sec of storage time.  The energy
scale is deduced from the time of arrival of the particles after the
release, with high energy particles escaping first.  Due to the
capacitance of the trap electrodes the relation between well depth and
release time is not linear and has been obtained by digitizing the exit
electrode potential vs. time.  We find  the width of the peak to be
less than $800$ meV, with the centroid located below $1$ eV. Due to
unknown contact potentials on the trap structure it is impossible to
determine the absolute value of the energy, and the width of the
distribution must be attributed mostly to the Coulomb interactions
amongst the charged particles (electrons and antiprotons) during their
release. Therefore, our results are fully compatible with $65$ \% of
the antiprotons having been cooled to the ambient temperature of the
trap ($< 15$ K) after $600$ seconds.

During the entire time between the initial capture of the antiproton
pulse and the final release from the inner trap, the counts in the
external scintillators are recorded.  Scintillators S1 - S3 are located
closest to the center of the trap and are therefore mostly sensitive to
annihilations occurring on the residual gas in the trap (see Figure
1).  For background suppression these scintillators are connected in a
two-fold coincidence set-up and the detection efficiency is determined
to be $4$ \%.  Since the number of stored antiprotons may vary in time,
the observed annihilation rate needs to be normalized to the number of
particles present in the trap at any given time $t$.  Such a normalized
annihilation rate, for a specific run, is shown in  Figure 4.

At the beginning of the cool-down we see an increase in the probability
for annihilation on the residual gas.  The annihilation rate reaches a
maximum at approximately $150$ seconds, but afterwards decreases
strikingly.  At $t = 600$ seconds the long, non-harmonic section of the
trap is opened and a  small, but sharp, drop in the annihilation rate
is seen.  This indicates the ejection of the few higher-energy
antiprotons  remaining in this section of the trap. Subsequently, the
observed rate is not distinguishable above the cosmic-ray background.
This is so even though, in this specific example, approximately $12$ \%
of the initially captured antiprotons were determined to be still
present in the inner trap at $t=1500$ sec.

This observation is in contradiction to the generally held belief that
the annihilation cross section at low energy should exhibit a $1/v$
dependence \cite{bracci}.  (This adiabatic calculation was done, it
should be noted, for hydrogen targets.) Such a $1/v$ behavior would
result in a normalized annihilation rate which would be independent of
the antiproton energy, which in turn implies that we should observe a
constant rate vs.  time.  Thus, with a $1/v$ behaviour, neither the
initial rise of the observed rate nor the decay at times larger than
200 seconds could be explained.  (The initial increase may be
consistent with a $1/{v^n}$, $n>1$, dependence as given, for instance,
by Morgan and Hughes \cite{morgan}, who had $n=2$.) The upper bound of
the observed annihilation rate is $8 \times 10^{-3}$ sec$^{-1}$.
However, the final annihilation rate at $t = 1500$ sec is significantly
lower than this.  To our knowledge no theoretical model exists that
predicts such a striking (or indeed, any) decrease of the rate with
temperature.  An approach \cite{voronin} different than that of Ref.
\cite{bracci} uses a coupled-channel, non-adiabatic procedure.
Although this produces a low rate at low energies, that model
underpredicts our measured results at times less than 200 sec.

The chemical composition of the residual gas in the trap is of critical
importance.  Since the outer wall of the vacuum vessel is in direct
contact with the liquid helium in the cryostat,  all gases except
hydrogen and helium should be frozen out.  Furthurmore, because of the
liquid helium environment and the fact that helium is poorly pumped by
the external ion-getter pumps, the remaining gas should be
predominantly helium.

Now we can consider the actual gas pressure in the trap.  In fact,  the
observed maximum annihilation rate, $8 \times 10^{-3}$ sec$^{-1}$, can
be used to verify a rough estimate for the residual gas pressure.
Assume that  (1) the cross-section estimates given by Bracci et al.
\cite {bracci} (which has a $1/v$ dependence) are valid as an upper
bound for our observed annihilation rate even through the calculation
was for a hydrogen target, and (2) the measured temperature of the trap
structure, 10 K, is the temperature of the residual gas.  Using these
parameters one obtains  $4 \times 10^{-12}$ Torr for the residual gas
pressure.  This  is in good agreement with our expectation that the gas
pressure in the trap is bounded from above by the lower limit of the
residual gas pressure in the cryogenic section, $10^{-11}$ Torr.

In those runs where no electrons were preloaded so no cooling of the
antiprotons was taking place, the observed annihilation rate was
constant over comparable time intervals. This shows the importance of
the temperature of the antiprotons and proves the stability of the
antiproton cloud against dynamical effects.

Experimental data for the annihilation of antiprotons on neutral
particles at low energy does not exist.  We are investigating the
possibility that there exists a small repulsive potential at short
range \cite{kinosh}.  Strong binding/antibinding effects on antiprotons
penetrating the electron cloud of helium atoms have been observed by
the PS194 collaboration \cite {PS194} in a study of the
double-ionization cross section for antiprotons and protons impacting
on a helium gas target at energies of 13 keV and above.  Recently, the
formation of metastable systems in antiproton-helium collisions have
been observed \cite{yamazaki} and theoretical predictions of repulsive
potentials in excited-state systems have been discussed \cite{briggs}.
Possibly related effects have been seen in positronium formation from
positron impact on large molecules \cite{surko}.  ( Elsewhere we will
comment in more detail on these points \cite{deutch}.)

The observed reduction of the annihilation rate at ultra-low energies
would have a significant impact on a number of experiments planned with
cold antiprotons. For these experiments antiprotons, once captured and
cooled in the PS200 catching trap, need to be extracted as a beam and
transported to either a scattering chamber or a second trap system for
recapture. Such transport would be made technically much easier if a
room temperature vacuum system can be used instead of enclosing the
entire apparatus in a cryogenic environment.

The construction of portable trap systems has also been proposed
\cite{portable}.  Antiprotons could then be delivered to laboratories
around the world, allowing many different kinds  of experiments to be
done. Such experiments could vary from ultra- low-energy antiproton
physics {\it per se} to scattering of several hundred MeV/c pions and
kaons (produced by low-energy antiproton annihilation on a production
target).

Portable traps will have to include a vacuum section which  can be
coupled first to the PS 200 catching trap (or a similar system) for
filling and which  can also be coupled to an experiment at a remote
site.  Again, this is easier to do if ultra-low pressures are not
needed. To summarize,  because of the reduced antiproton annihilation
rate at low energies that we have observed, the long storage times
needed for both transport of and also experimentation with antiprotons
can realistically be achieved.

Future work will include the controlled reheating of the cooled
antiprotons.  This will be done by using resonance excitation of the
axial motion with radio-frequency fields.  The energy dependence of the
annihilation cross section will be studied.  We also will use different
target gases to investigate the possible effect of the polarization
potential of the target atom.

The work described here has been performed within the framework of the
PS200 experimental development and we wish to thank the entire PS200
collaboration for their support. We especially wish to thank P. L. Dyer
for the development of the data acquisition system used for these
measurements, J. Rochet for his assistance in constructing and
operating the experimental apparatus, and M. Charlton and Y. Yamazaki
for their support during data taking.  We appreciate the helpful
comments by S. Barlow on the positron annihilation data. None of the
results presented here would have been obtainable without the support
and help from the entire LEAR operating team. A very special `thank
you' goes to J.-Y. Hemery, M. Michel, and M. Giovannozzi for delivering
the very best beam spot possible to the entrance of our apparatus.
This work was in part supported by the US Department of Energy under
contract no. W7405 ENG-36 (Los Alamos), the U.S. Air Force Office of
Scientific Research under grant no. F49620-94-1-0223 (Pennsylvania
State University) and the Alexander von Humboldt Stiftung (M.M.N. at
Ulm).

\newpage

Figure Captions:

\begin{description}

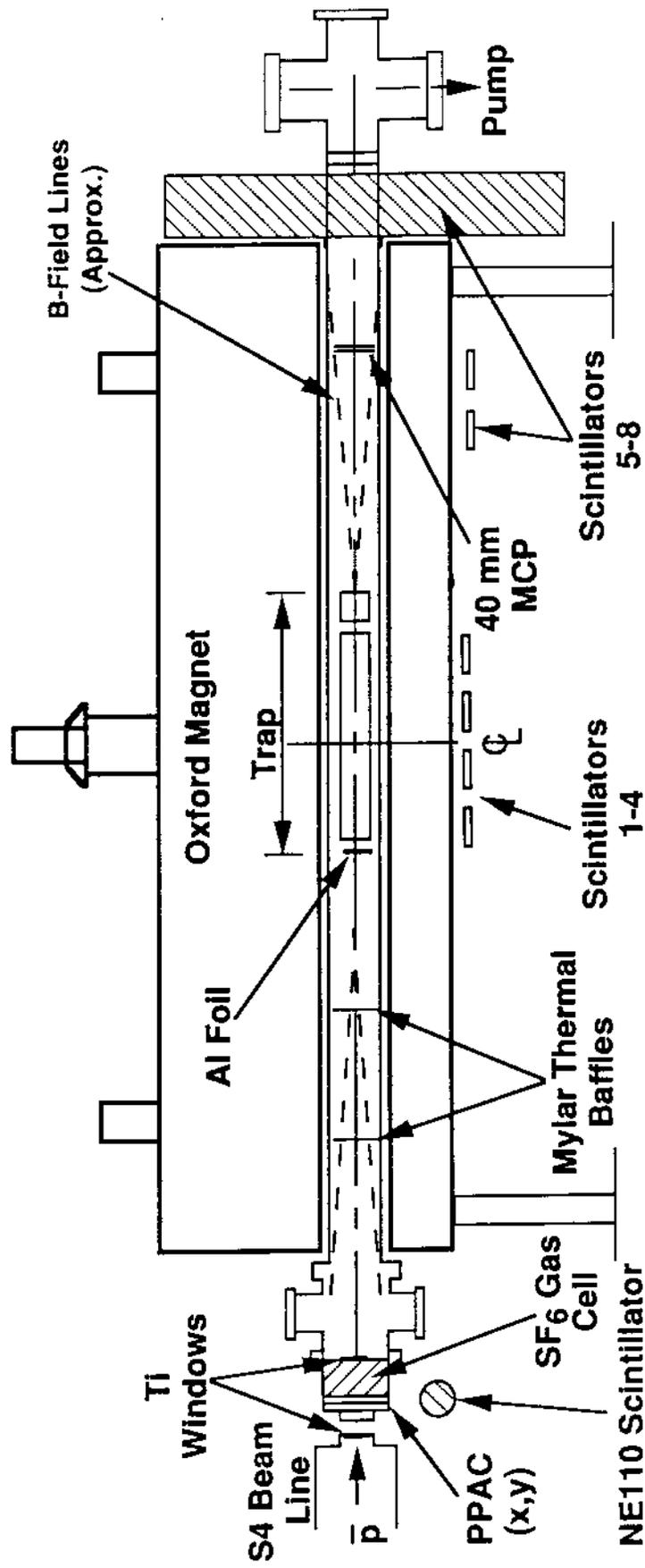
\item[Fig.1] {Schematic layout of the experimental set-up. Shown is the
superconducting magnet system (length 2 meter), the PS200 catching
trap, all beam monitors, and the scintillators used to trigger the
voltage switch and to monitor the  antiproton annihilations during
storage and upon release.}

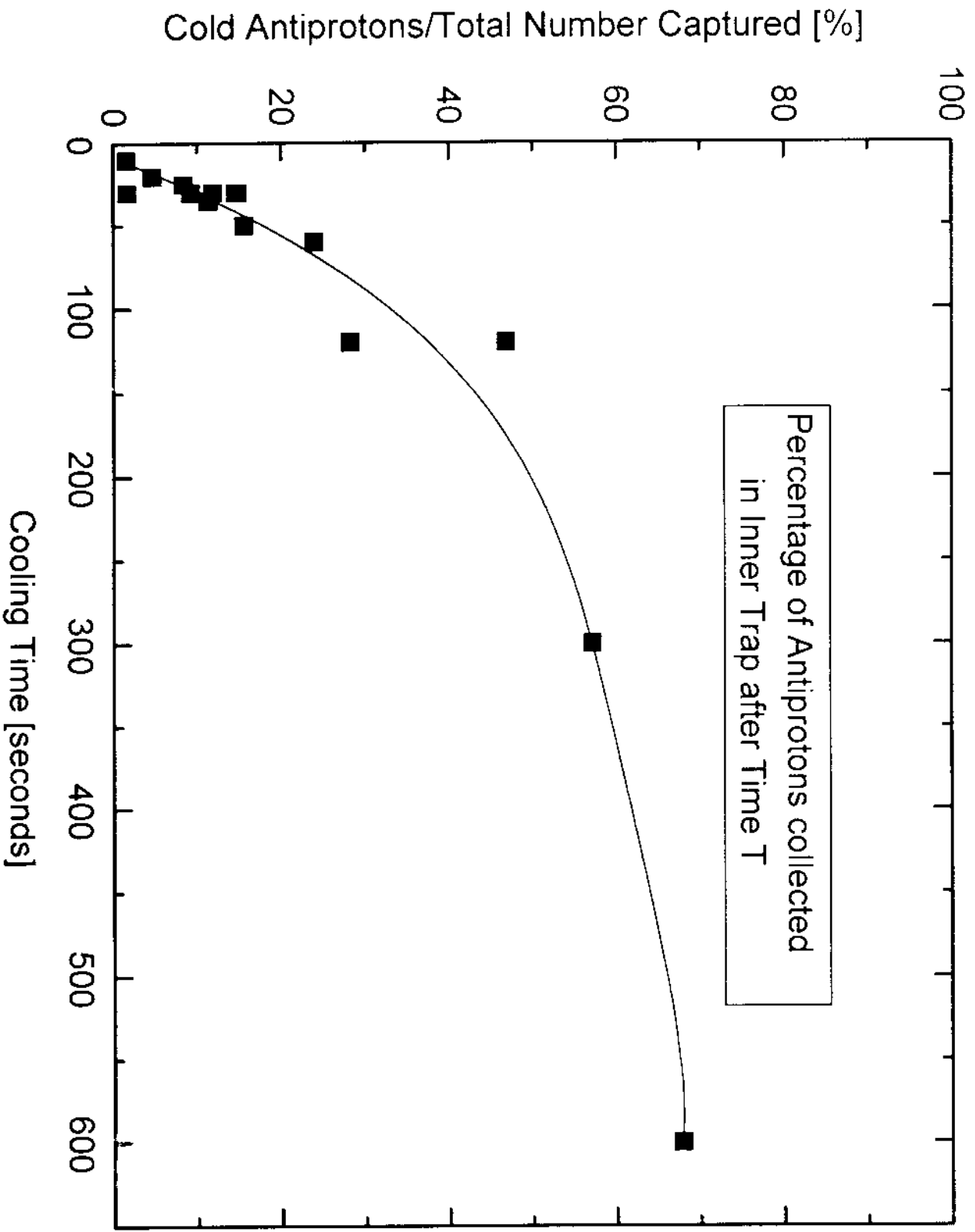
\item[Fig.2] {Accumulation of ultra-low energy antiprotons in the
harmonic well in the center of the PS200 catching trap. The solid line
is calculated for a cooling time constant of $175$ seconds and a
maximum transfer efficiency of $65$ \%.}

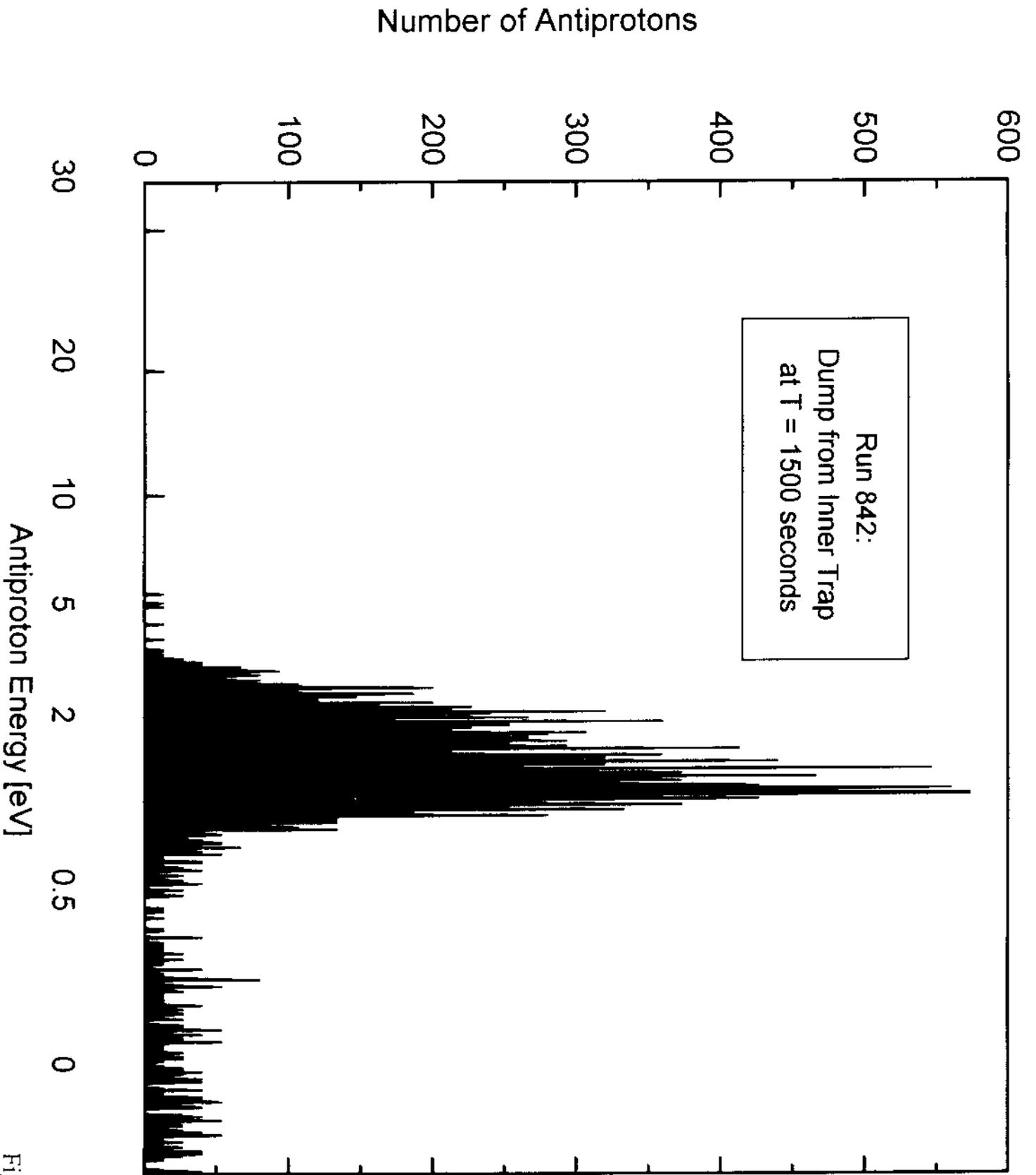
\item[Fig.3] {Energy spectrum of cold antiprotons released from the
inner trap.  Note that the energy scale is in the reverse direction and
is quite nonlinear towards the low-energy end.  The centroid of the
distribution is at $\leq 1$ eV, the FWHM is $\leq 800$ meV.}

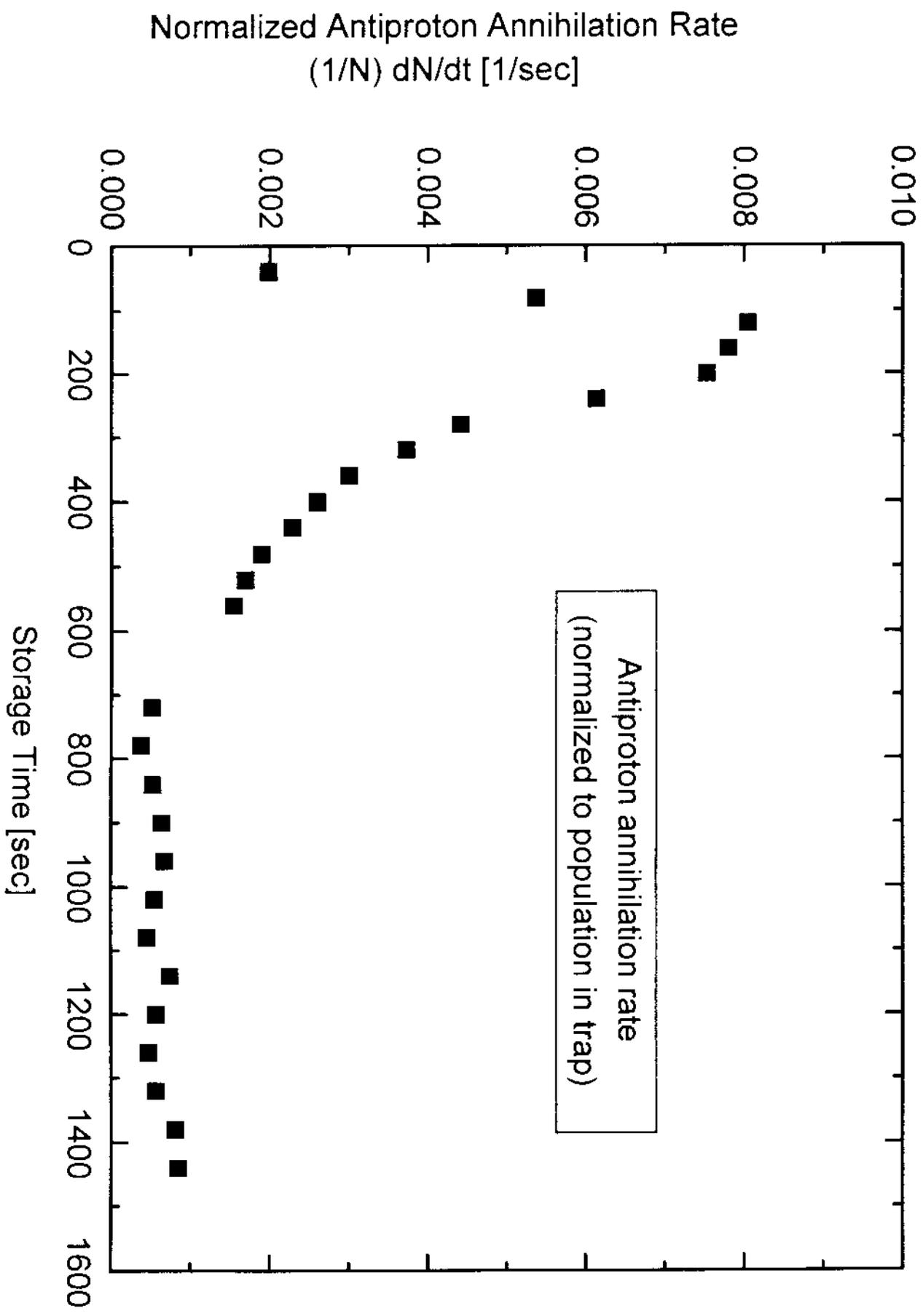
\item[Fig.4] {Rate of annihilation during storage and cooling of
antiprotons in the PS200 catching trap. The observed rate has been
normalized to the number of antiprotons in the trap at any given time
$t$. The sharp drop between $600$ and $700$ seconds is due to the loss
of antiprotons when the outer trap is opened completely.}

\end{description}

\end{document}